\documentclass[vecphys]{svmult}

\usepackage{makeidx}         
\usepackage{graphicx}        
\usepackage{multicol}        
\usepackage[bottom]{footmisc}

\makeindex             

\begin{document}
\title*{Maximum brightness temperature for an incoherent synchrotron radio source}
\titlerunning{Maximum brightness temperature}
\author{Ashok K. Singal}

\institute{Astronomy \& Astrophysics Division, Physical Research Laboratory,
Navrangpura, Ahmedabad, India - 380009.
\texttt{asingal@prl.res.in}}

\maketitle
\begin{abstract}
  We discuss here a limit on the maximum brightness temperature
  achievable for an incoherent synchrotron radio source. This limit,
  commonly referred to in the literature as an inverse Compton limit,
  prescribes that the brightness temperature for an incoherent
  synchrotron radio source may not exceed $\sim 10^{12} $~K, a fact
  known from observations. However one gets a somewhat tighter limit
  on the brightness temperatures, $T_{\rm b}\stackrel{<}{_{\sim}} 10^{11.5} $~K,
  independent of the inverse Compton
  effects, if one employs the condition of equipartition of energy in
  magnetic fields and relativistic particles in a synchrotron radio
  source. Pros and cons of the two brightness temperature limits are
  discussed.
\end{abstract}
\section{Introduction}
\label{sec:1}

From the radio spectra combined with the VLBI (Very large Baseline
Interferometery) observations it has been seen that the brightness
temperatures for compact self-absorbed radio source do not exceed
about $\sim 10^{11-12} $~K. Kellermann and Pauliny-Toth \cite{AKS1} first
time gave an explanation of this in terms of what has since then come
to be known in the literature as an Inverse Compton limit. They argued
that at brightness temperature $T_{\rm b} \stackrel{>}{_{\sim}}
10^{12} $~K energy losses of radiating electrons due to inverse
Compton effects become so large that these result in a rapid cooling
of the system, thereby bringing the synchrotron brightness temperature
quickly below this limit. Of course much larger brightness
temperatures have been inferred for the variable sources, however this
excess in brightness temperatures has been explained in terms of a
bulk relativistic motion of the emitting component \cite{AKS2,AKS3,
AKS4}. The Doppler factors required to explain the excess in
temperatures were initially thought to be $\sim 5 - 10$, similar to
the ones required for explaining the superluminal velocities seen in
some compact radio sources \cite{AKS5,AKS6,AKS7}.  
Singal and Gopal-krishna \cite{AKS8} pointed out that 
under the conditions of equipartition of energy between magnetic fields and 
relativistic particles in a synchrotron radio source, much higher Doppler 
factors are needed to successfully explain the variability events. 
Singal \cite{AKS9}, without taking recourse to any inverse 
Compton effects, derived a somewhat tighter upper limit 
$T_{\rm b}\stackrel{<}{_{\sim}} 10^{11.5} $~K, first time using the 
argument that due to the diamagnetic effects the energy
in the magnetic fields cannot be less than a certain fraction of that
in the relativistic particles and then an upper limit on brightness 
temperature follows naturally. 
However, it has to be noted that if one considers the drift currents at 
the boundaries, which may be present due to the 
non-uniformities there in the magnetic fields \cite{AKS10}, 
the above limit on the magnetic field energy gets modified. 
Later similar derivations \cite{AKS11} of the 
$T_{\rm  b}\stackrel{<}{_{\sim}} 10^{11.5}$~K limit 
as well as of large Doppler factors, used essentially the same 
equipartition arguments as in \cite{AKS8,AKS9}.

Here we first derive the inverse Compton limit on $T_{\rm b}$ using the 
approach followed in \cite{AKS1}, and then  the 
equipartition limit as done in \cite{AKS9}. In fact we argue that 
even if we relax the condition of the equipartition of energy between magnetic 
fields and relativistic particles and let the energy in relativistic 
particles to be many orders of magnitude larger than that in the magnetic 
fields we still end up with a rather tight $T_{\rm b}$ limit. On the other 
hand if the energy in particles is {\em smaller} than that in magnetic fields, 
then in any case $T_{\rm  b}$ {\em has to be lower than} 
$\sim 10^{11.5} $~K, as pointed out in \cite{AKS9}.

\section{Inverse Compton limit}
\label{sec:2}

We want to study the maximum brightness temperature limit in the rest
frame of the source, therefore we assume that all quantities have been
transformed to that frame. Hence we will not consider here any effects
of the cosmological redshift or of the relativistic beaming due to a
bulk motion of the radio source.

In inverse Compton interaction between relativistic electrons and
their synchrotron photons, the energy of radio photons could get
boosted to X-ray frequencies. The average energy of a photon in an
inverse Compton interaction gets boosted by a factor $\gamma_{e}^{2}$
where $\gamma_{e}$ is the Lorentz factor of the interacting
relativistic electrons \cite{AKS12}.  We confine our
discussion to a single scattering case only.

A relativistic electron of Lorentz factor $\gamma_{e}$ 
gyrating in a magnetic field $B$ emits most of its radiation 
in a frequency band near its characteristic synchrotron frequency 
\cite{AKS13,AKS14} 
\begin{equation}
\nu_{c} = 0.29 \frac {3}{4 \pi}\frac{e\, B}{m_{0}\, c}\gamma_{e}^{2} 
\end{equation}
which gives us $\gamma_{e}^{2}=8.2 \times 10^{2}\, \nu_{c}/B$ for
$\nu_{c}$ in GHz and $B$ in Gauss. From this we find that for $B\sim
10^{-3}$ Gauss as inferred in compact radio sources, $\nu_{c} = 0.1 $
to 10 GHz yields $\gamma_{e}^{2} \sim 10^{5} $ to $\sim 10^{7} $. Thus
during an inverse Compton interaction while at the lower end the
synchrotron photons can get boosted to infrared frequencies ($\sim
10^{13}$) Hz, at the higher end the synchrotron photons of radio
frequencies $\sim 10^{10}$ Hz could get boosted to X-ray frequencies
($\sim 10^{17}$) Hz.

The power radiated by the inverse Compton process $P_{\rm c}$ as
compared to that radiated by synchrotron mechanism, $P_{\rm s}$ is
given as,
\begin{equation}
\frac {P_{\rm c}}{P_{\rm s}}= \frac {W_{\rm p}}{W_{\rm b}}
\end{equation}
where $W_{\rm p}$ and $W_{\rm b}$ respectively are the photon energy
density and magnetic field energy density within the source. This
relation is true in the case where Thomson scattering cross-section is
valid \cite{AKS12}, which is true in our case as we
have $\gamma_{e} h \nu << m_{0} c^{2}$, for $\nu < 100$ GHz.

The photon energy density is related to the radiation intensity as
\cite{AKS14}
\begin{equation}
W_{p}=\frac{4 \pi}{c} \int_{\nu_{1}}^{\nu_{2}}I_{\nu}\: {\rm d}\nu
\end{equation}
where the specific intensity $I_{\nu}$ is defined as the flux density
per unit solid angle, at frequency $\nu$. In radio sources, the
observed flux density in the optically thin part of the spectrum
usually follows a power law, i.e., $I_{\nu} \propto \nu^{-\alpha} $,
between the lower and upper cut off frequencies $\nu_{1}$ and
$\nu_{2}$. In synchrotron theory this spectrum results from a power
law energy distribution of radiating electrons $N(E)\propto
E^{-\gamma}$ within some range $E_{1}$ and $E_{2}$, with $\gamma =
2\alpha+1$ and $E_{1}$, $E_{2}$ related to $\nu_{1}$, $\nu_{2}$ by
Eq. (1).  In compact radio sources the source may become self-absorbed
with flux density $\propto \nu^{2.5}$ below a turnover frequency
$\nu_{\rm m}$.  From Eq.(3) we can get
\begin{equation}
W_{p}= 2.27 \times 10^{-7} \:  \frac{f(\alpha)\: F_{\rm m} \, \nu_{m}^{\alpha}}
{\Theta_{\rm x}\: \Theta_{\rm y}}  \left[
\frac{\nu_{2}^{1-\alpha} - \nu_{\rm 1}^{1-\alpha}}{1-\alpha} \right] {\rm erg\: cm}^{-3}
\end{equation}
the expression to be evaluated in the `limit' for $\alpha=1$.  Here
$F_{\rm m}$ (Jy) is the flux density at frequency $\nu_{\rm m}$ (GHz)
corresponding to the point of spectral turnover, while $\Theta_{\rm
  x}$ and $\Theta_{\rm y}$ (mas) represent the angular size along the
major and minor radio axes of the source component, assumed to be an
ellipse. For calculating the photon energy density it may be
appropriate to take the lower limit of the spectrum at $\nu_{\rm m}$
itself, therefore we can put $\nu_{1}=\nu_{\rm m}$ in the above
expression. Here it should be noted that $F_{\nu}$ is the flux density
in the optically thin part of the synchrotron spectrum, accordingly
$F_{\nu_{\rm m}}$ obtained from an extrapolation up to $\nu = \nu_{m}$
using the straight slope $\alpha$, is not the same as the actual peak
flux density $F_{\rm m}$ at the turnover bend (see e.g., \cite{AKS15}). 
In Table 1 values of $f(\alpha)=F_{\nu_{\rm m}}/F_{\rm
  m}$ are listed, which as we see are of the order of unity.

We can express $W_{p}$ in terms of the brightness temperature at the
peak of the spectrum
\begin{equation}
  T_{\rm m} = 1.763 \times 10^{12} F_{\rm m}\, \Theta_{\rm x}^{-1} \Theta_{\rm y}^{-1}  
  \nu_{\rm m}^{-2}\,  {\rm K}
\end{equation}
to get
\begin{equation}
W_{p}= 1.29 \times 10^{-7} 
\: \frac{f(\alpha)}{1-\alpha} \left[
\left(\frac{\nu_{2}}{\nu_{\rm m}}\right)^{1-\alpha} - 1 \right] \,  \nu_{\rm m}^{3}
\left(\frac{T_{\rm m}}{10^{12}}\right) {\rm erg\: cm}^{-3}. 
\end{equation}

On the other hand, from the synchrotron self-absorption we get,
\begin{equation}
B =  10^{-5}\, b(\alpha)\, F_{\rm m}^{-2}\, \Theta_{\rm x}^{2}\, \Theta_{\rm y}^{2}\,
 \nu_{\rm m}^{5},
\end{equation}
values of $b(\alpha)$\footnote{Calculated from the tabulated functions in 
\cite{AKS14}.} are given in Table 1.

Here a plausible assumption has been made that the direction of the
magnetic field vector, with respect to the line of sight, changes
randomly over regions small compared to a unit optical depth. For a
uniform magnetic field direction throughout the source region, the
co-efficients $a(\alpha)$ (see next section) and $b(\alpha)$ would be
modified by factors of the order of unity.

Then the magnetic field energy density $W_{b}= B^{2}/8 \pi$ can be
written as,
 
\begin{equation}
W_{b}= 3.84 \times 10^{-11} \, b^{2}(\alpha)\,\nu_{\rm m}^{2} \left( \frac
{T_{\rm m}}{10^{12}}\right)^{-4} \,{\rm erg\: cm}^{-3}.
\end{equation} 
Equations (2), (6) and (8) lead us to,
\begin{equation}
\frac {P_{\rm c}}{P_{\rm s}}= \nu_{\rm m} 
\left(\frac{T_{\rm m}}{10^{11.3} \: p(\alpha)} \right)^{5} 
\end{equation}
where the function
\begin{equation} 
p(\alpha)=\left[\frac{f(\alpha)}{b^{2}(\alpha)(1-\alpha)} \left\{
\left(\frac{\nu_{2}}{\nu_{\rm m}}\right)^{1-\alpha} - 1 \right\} \right]^{-1/5}
\end{equation} 
is of the order of unity (Table 1)\footnote{In Table 1 and 2 we have
  taken $\nu_{1}$ and $\nu_{2}$ to be 0.01 and 100 GHz, and the
  turnover frequency $\nu_{\rm m}$ is taken to be 1 GHz.}.

\section{Equipartition temperature limit}

\label{sec:3}

Energy density of the relativistic electrons in a synchrotron radio
source component is given by \cite{AKS13,AKS16}
\footnote{Values of $a(\alpha)$ in Table 1, calculated by us from 
the tabulated functions given in \cite{AKS14}, appear slightly different from 
the ones in \cite{AKS13} but are in agreement with those in \cite{AKS17}.}  

\begin{equation}
W_{\rm e}= \frac {8.22 \times 10^{-9}} {a(\alpha)\:(\alpha-0.5)} 
\frac {F_{\nu}\: \nu^{\alpha}\: B^{-1.5}} {\Theta_{\rm x}\: \Theta_{\rm y}\: s}
\left[\left(\frac {y_{1}(\alpha)}{\nu_{1}}\right) ^{\alpha-0.5} 
-\left(\frac {y_{2}(\alpha)}{\nu_{2}}\right) 
^{\alpha-0.5}   \right]
\end{equation}
this expression to be evaluated in the `limit' for $\alpha=0.5$. 
Here $W_{\rm e}$ (erg cm$^{-3}$) is the energy density of radiating electrons;
$F_{\nu}$ (Jy) is the flux density at frequency $\nu$ (GHz), 
with  $\nu_{1} < \nu < \nu_{2}$; and $s$(pc) is the characteristic depth of 
the component along the line of sight. 
%
\begin{table}
\centering
\caption{Various functions of the spectral index $\alpha$}
\label{tab:1}       
%
%
\begin{tabular}{lclccllcl}
\hline\noalign{\smallskip}
$\;\alpha \;$  &  $\;\;\gamma \;\;\;$ & $\; a(\alpha)$ & $\; b(\alpha)$ 
& $\;f(\alpha)\;$ &  $p(\alpha)\;$ &  $t(\alpha)\;$ & $y_{1}(\alpha)$ & $y_{2}(\alpha)$ \\
\noalign{\smallskip}\hline\noalign{\smallskip}
0.25 & 1.5 & 0.149  & 2.07 & 1.10  & 0.66  & 0.68  & 1.3 & 0.011 \\
0.5  & 2.0 & 0.103  & 2.91 & 1.19  & 0.83  & 0.85  & 1.8 & 0.032 \\
0.75 & 2.5 & 0.0831 & 2.85 & 1.27  & 0.94  & 0.80  & 2.2 & 0.10  \\
1.0  & 3.0 & 0.0741 & 2.52 & 1.35  & 1.00  & 0.67  & 2.7 & 0.18  \\
1.5  & 4.0 & 0.0726 & 1.79 & 1.50  & 1.03  & 0.43  & 3.4 & 0.38  \\
\noalign{\smallskip}\hline
\end{tabular}
\end{table}

In terms of the brightness temperature we can write,
\begin{equation}
W_{\rm e}= \frac {4.66 \times 10^{-9}\: f(\alpha)} {s\: a(\alpha)\:(\alpha-0.5)}
\left[\left(\frac {y_{1}(\alpha)}{\nu_{1}/\nu_{\rm m}}\right) ^{\alpha-0.5} 
-\left(\frac {y_{2}(\alpha)}{\nu_{2}/\nu_{\rm m}}\right)^{\alpha-0.5} \right]
\frac{\nu_{\rm m}^{2.5}} {B^{1.5}}\left(\frac{T_{\rm m}}{10^{12}}\right).
\end{equation}

From the overall charge neutrality of the plasma we expect the
electrons to be accompanied by an equal number of positive
charges. Any positrons would have already been accounted for in our
equation above, however presence of heavy particles will contribute
additionally to the total particle energy density. Let the energy in
the heavy particles be $\xi$ times that in the lighter particles, then
the total particle energy density is $W_{\rm k}=(1+\xi)\: W_{\rm e}$.

Now if we assume the energy in particles to be related to that in
magnetic fields by
\begin{equation}
W_{\rm k}= \eta \:W_{\rm b}
\end{equation}
then we get
\begin{eqnarray}
B^{7/2}=\frac {1.17 \times 10^{-7}f(\alpha)}{a(\alpha)\: (\alpha-0.5)}  
\left[\left(\frac {y_{1}(\alpha)}{\nu_{1}/\nu_{\rm m}}\right) ^{\alpha-0.5} 
-\left(\frac {y_{2}(\alpha)}{\nu_{2}/\nu_{\rm m}}\right) ^{\alpha-0.5} \right]\;\;\times 
\;\;\;\;\;\;\;\;\;\;\;\;\; 
\nonumber \\
\frac {(1+\xi)}{s\: \eta}\;\nu_{\rm m}^{2.5} \left(\frac{T_{\rm m}}{10^{12}}\right). 
\;\;\;\;\;
\end{eqnarray}
Now substituting for magnetic field from Eq.(7) we get,
\begin{equation}
\frac {\eta}{1+\xi} = \frac{1}{ s \,\nu_{\rm m}}
\left(\frac{T_{\rm m}}{10^{10.9}\: t(\alpha)} \right)^{8} 
\end{equation}
where the function  
\begin{equation}
t(\alpha)= \left[\frac {f(\alpha)}{(\alpha-0.5)\; a(\alpha)\; b^{3.5}(\alpha)}\left \{
\left(\frac {y_{1}(\alpha)}{\nu_{1}/\nu_{\rm m}}\right) ^{\alpha-0.5} -
\left(\frac {y_{2}(\alpha)}{\nu_{2}/\nu_{\rm m}}\right) ^{\alpha-0.5}   \right \} 
\right] ^{-1/8} 
\end{equation}
is of the order of unity (Table 1).

\section{A Correction to the derived $T_{\rm m}$ values}
\label{sec:4}

Actually $T_{\rm m}$ values have been calculated (both here as well as
in the literature) for the turnover point in the synchrotron spectrum
where the flux density peaks. However, the definition of the
brightness temperature (Eq. 5) also involves $\nu^{-2}$. Therefore a
maxima of flux density is not necessarily a maxima for the brightness
temperature also. In fact a zero slope for the flux density with
respect to $\nu$ would imply for the brightness temperature a slope of
$-2$. Therefore the peak of the brightness temperature will be at a
point where flux density $\propto \nu^{2}$, so that $T_{b} \propto F
\, \nu^{-2}$ has a zero slope with respect to $\nu$.  The peak of
$F_{\nu} \, \nu^{-2}$, can be determined in the following manner.  The
specific intensity in a synchrotron self-absorbed source is given by
\cite{AKS14},
\begin{equation}
I_{\nu} = \frac {c_{5}(\alpha)}{c_{6}(\alpha)} \left(\frac {\nu}{2c_{1}}\right)^{2.5} 
 \left[1-\exp\left\{-\left( \frac{\nu}{\nu_{1}}\right)^{-(\alpha+2.5)}\right\}\right] 
\:B_{\perp}^{-0.5}
\end{equation}
where $c_{1}, c_{5}(\alpha), c_{6}(\alpha)$ are tabulated in \cite{AKS14}. 
The optical depth varies with frequency as $\tau=(\nu/\nu_{1})^{-(\alpha+2.5)}$,  
$\nu_{1}$ being the frequency  at which $\tau$ is unity. The equivalent brightness 
temperature (in Rayleigh-Jeans limit) is then given by
\begin{equation}
T_{\nu} = \frac {c^{2}\, c_{5}(\alpha)}{8\, k\, c_{1}^{2}\, c_{6}(\alpha)} 
\left(\frac {\nu}{2c_{1}}\right)^{0.5} 
 \left[1-\exp\left\{-\left( \frac{\nu}{\nu_{1}}\right)^{-(\alpha+2.5)}\right\}\right] 
\:B_{\perp}^{-0.5}
\end{equation}
where $k$ is the Boltzmann constant and $c$ is the speed of light.  We
can maximize $T_{\nu}$ by differentiating it with $\nu$ and equating
the result to zero. This way we get an equation for the optical depth
$\tau_{\rm o}$, corresponding to the {\em peak brightness temperature}
$T_{\rm o}$, which is different from the one that is available in the
literature for the optical depth $\tau_{\rm m}$ at the {\em peak of
  the spectrum}.  The equation that we get for $\tau_{\rm o}$ is
\begin{equation} 
\exp\,(\tau_{\rm o})=1+(2\alpha+5)\,\tau_{\rm o},
\end{equation}
solutions of this transcendental equation for different $\alpha$
values are given in Table 2. It is interesting to note that while the
peak of the spectrum for the typical $\alpha$ values usually lies in
the optically thin part of the spectrum ($\tau_{\rm
  m}\stackrel{<}{_{\sim}}1$; Table 2), peak of the brightness
temperature lies deep within the optically thick region ($\tau_{\rm o}
\sim 3$). Both the frequency and the intensity have to be calculated for
$\tau_{\rm o}$ to get the maximum brightness temperature values. The
correction factors are then given by,
\begin{equation}
\frac{\nu_{\rm o}}{\nu_{\rm m}}=\left(\frac{\tau_{\rm m}}{\tau_{\rm o}}\right)^{1/(\alpha+2.5)}
\end{equation}
\begin{equation}
\frac{T_{\rm o}}{T_{\rm m}}=\left(\frac{\nu_{\rm o}}{\nu_{\rm m}}\right)^{0.5}\,
\left[\frac {1-\exp\,(-\tau_{\rm o})}{1-\exp\,(-\tau_{\rm m})}\right]
\end{equation}

\begin{table}
\centering
\caption{Values for the temperature limits}
\label{tab:2}       
%
%
\begin{tabular}{lcllcccc}
\hline\noalign{\smallskip}
$\;\alpha \;$  &  $\;\;\gamma \;\;\;$ & $\;\tau_{\rm m}\;\;$ 
& $\;\tau_{\rm o}$  & $\nu_{\rm o}/\nu_{\rm m}$ & $\;T_{\rm o}/T_{\rm m}$ & 
$\log\,(T_{\rm ic})$ & $\log\,(T_{\rm eq})$\\
\noalign{\smallskip}\hline\noalign{\smallskip}
0.25 & 1.5 & 0.19 & 2.80 & 0.37 & 3.5 & 11.7 & 11.3 \\
0.5  & 2.0 & 0.35 & 2.92 & 0.50 & 2.2 & 11.6 & 11.2 \\
0.75 & 2.5 & 0.50 & 3.03 & 0.58 & 1.8 & 11.5 & 11.1 \\
1.0  & 3.0 & 0.64 & 3.13 & 0.64 & 1.6 & 11.5 & 10.9 \\
1.5  & 4.0 & 0.88 & 3.32 & 0.72 & 1.4 & 11.5 & 10.7 \\
\noalign{\smallskip}\hline
\end{tabular}
\end{table}
In Table 2 we have listed $\nu_{\rm o}/\nu_{\rm m}$ and  $T_{\rm o}/T_{\rm m}$, 
for different $\alpha$ values. Further we have also
tabulated values of maximum brightness temperatures $\log\,(T_{\rm ic})$ for 
$W_{\rm p}= W_{\rm b}$ and $\log\,(T_{\rm eq})$ for $\eta=1$ (with $\xi=0$, $s=1$pc) 
calculated from equations (9) and (15) respectively for 
different $\alpha$ values, incorporating the above corrections.

\section{Discussion and Conclusions}
\label{sec:5}
From Table 2 we see that, depending upon the spectral index value
$\alpha$, theoretical maximum brightness temperature values for the
inverse Compton limits are around $T_{\rm ic} \sim
10^{11.6\stackrel{+}{\_}0.1} $~K while for equipartition conditions
the limits are lower by a factor of $\sim 3$ ($T_{\rm eq} \sim
10^{11.0\stackrel{+}{\_}0.3} $~K).  From the observational data 
\cite{AKS18} it seems that the intrinsic $T_{\rm b}$ is
$\stackrel{<}{_{\sim}} 10^{11.3} $~K, which is broadly consistent with
the either interpretation. It should be noted that though inverse
Compton scattering increases the photon energy density, yet it does
not increase the radio brightness as the scattered photons get boosted
to much higher frequency bands. If anything, some photons get removed
from the radio window, but the change in radio brightness due to that
may not be very large.  What could be important is the large energy
losses by electrons which may cool the system rapidly. However, these
inverse Compton losses become important only when $T_{\rm b} >
10^{11.5} $~K, but the equipartition conditions may keep the
temperatures well below this limit. This is not to say that inverse
Compton effects cannot occur, it is only that conditions in
synchrotron radio sources may not arise for inverse Compton losses to
become very effective. Since we are considering the brightness
temperature limit in the radio-band (after all that is where
observationally such limits have been seen), then variations in
$\nu_{\rm m}$ that we may consider would at most be about an
order of magnitude around say, 1 GHz.  With the reasonable assumption
that a self-absorbed radio source size may not be much larger than
$\sim$ a pc, from Eq. (15) it follows that an order of magnitude
higher $T_{\rm m}$ values would require $\eta$ to increase by about a
factor ${\sim} 10^{8}$, that is departure from equipartition will go
up by about eight orders of magnitude.  Actually for a given $\nu_{\rm
  m}$, the magnetic field energy density will go down by a factor
$\sim 10^{4}$ (Eq. 8), while that in the relativistic particles will
go up by a similar factor (Eq. 12, note the presence of $B^{1.5}$ in
the denominator).  This also implies that the total energy budget of
the source will be higher by about $\sim 10^{4}$ than from the already
stretched equipartition energy values. Further, it is not the photon
energy density that really goes up drastically with higher brightness
temperatures (as $W_{p} \propto T_{\rm m}$, Eq. 6), rather it is the
drastic fall in the magnetic field energy ($W_{b} \propto T_{\rm
  m}^{-4}$, Eq. 8) that increases the ratio $P_{\rm c}/P_{\rm s}$ to
go up as $T_{\rm m}^{5}$. Therefore, it is still not clear whether the
inverse Compton effects actually do play a significant role in
maintaining the maximum brightness temperature limit in incoherent
synchrotron radio sources.


\begin{thebibliography}{99.}
%
%
%
\bibitem{AKS1} K. I. Kellermann, I. I. K. Pauliny-Toth: ApJ \textbf{155}, L71 (1969)
\bibitem{AKS2} M. J. Rees: Nature \textbf{211}, 468 (1966)
\bibitem{AKS3} M. J. Rees: MNRAS \textbf{135}, 345 (1967)
\bibitem{AKS4}L. Woltjer: ApJ \textbf{146}, 597 (1966)
\bibitem{AKS5}M. H. Cohen, W. Cannon, G. H. Purcell et al: ApJ \textbf{170}, 207 (1971)
\bibitem{AKS6}A. R. Whitney, I. I., Shapiro, A. E. E., Rogers et al: 
Science \textbf{173}, 225 (1971)
\bibitem{AKS7}M. H. Cohen, R. P. Linfield, A. T. Moffet et al: Nature \textbf{268}, 405 (1977)
\bibitem{AKS8}A. K. Singal, Gopal-krishna: MNRAS \textbf{215}, 383 (1985)
\bibitem{AKS9}A. K. Singal: A\&A \textbf{155}, 242 (1986)
\bibitem{AKS10}G. Bodo, G. Ghisellini, E. Trussoni: MNRAS \textbf{255}, 694 (1992)
\bibitem{AKS11}A. C. S. Readhead: ApJ \textbf{426}, 51 (1994)
\bibitem{AKS12} G. B. Rybicki, A. P. Lightman: \textit{Radiative Processes in Astrophysics}, 
(Wiley, New York 1979)
\bibitem{AKS13}V. L. Ginzburg, S. I. Syrovatskii: ARAA \textbf{3}, 297 (1965)
\bibitem{AKS14} A. G. Pacholczyk: \textit{Radio Astrophysics}, (Freeman, San Francisco 1970) 
\bibitem{AKS15}M. A. Scott, A. C. S. Readhead: MNRAS \textbf{180}, 539 (1977) 
\bibitem{AKS16}V. L. Ginzburg:  \textit{Theoretical Physics and Astrophysics}, 
(Pergamon, Oxford 1979) 
\bibitem{AKS17}R. J. Gould: A\&A \textbf{76}, 306 (1979)
\bibitem{AKS18}D. C. Homan, Y. Y. Kovalev, M. L. Lister et al: ApJ \textbf{642}, L115 (2006)
\end{thebibliography}
%



\printindex
\end{document}